\documentclass[prb,twocolumn,superscriptaddress,showpacs,floatfix]{revtex4}
\usepackage{amssymb}
\usepackage{amsfonts}
\usepackage{subfigure}
\usepackage{graphicx}
\usepackage{dcolumn}
\usepackage{bm}
\usepackage{hyperref}
\usepackage[mathlines]{lineno}
\usepackage{amsmath}

\preprint{}

\begin{document}

\title{Dirac-Kondo semimetals and topological Kondo insulators in the dilute carrier limit}

\author{Xiao-Yong Feng}
  \affiliation{Condensed Matter Group,
  Department of Physics, Hangzhou Normal University, Hangzhou 310036, China}

\author{Hanting Zhong}
  \affiliation{Condensed Matter Group,
  Department of Physics, Hangzhou Normal University, Hangzhou 310036, China}

\author{Jianhui Dai}
  \affiliation{Condensed Matter Group,
  Department of Physics, Hangzhou Normal University, Hangzhou 310036, China}

\author{Qimiao Si}
  \affiliation{Department of Physics and Astronomy, Rice University,
Houston, TX 77005, USA }


\begin{abstract}
Heavy fermion systems contain not only strong electron correlations,
which promote a rich set of quantum phases, but also a large
spin-orbit coupling, which tends to endow the electronic states a
topological character. Kondo insulators are understood in terms of a
lattice of local moments coupled to conduction electrons in a
half-filled band,  {\it i.e.} with a dense population of about one
electron per unit cell. Here, we propose that a new class of Kondo
insulator arises when the conduction-electron band is nearly empty
(or, equivalently, full). We demonstrated the effect through a
honeycomb Anderson lattice model. In the empty carrier limit,
spin-orbit coupling produces a gap in the hybridized heavy fermion
band, thereby generating a topological Kondo insulator. This state
can be understood in terms of a nearby phase in the overall phase
diagram, a Dirac-Kondo semimetal whose quasiparticle excitations
exhibit a non-trivial Berry phase. Our results point to the dilute
carrier limit of the heavy-fermion systems as a new setting to study
strongly correlated insulating and topological states.
\end{abstract}

 \maketitle

\vskip 0.5 cm


Heavy fermion compounds represent a prototype system to study the
physics of strong electron correlations. These materials are
described in terms of $f$-electron-derived local moments and, in the
simplest case, a band of conduction electrons. At low temperatures,
the local moments and the spins of conduction electrons are
entangled into a spin-singlet state. This Kondo effect turns the
local moments into fermionic quasiparticles, which hybridize with
the conduction electrons to yield a metal whose effective carrier
mass is as large as thousands that of the bare electron mass. This
heavy-fermion metal phase has been the starting point to realize
quantum phase transitions into a variety of antiferromagnetically
ordered phases \cite{Coleman-Schofield05,Si-Steglich10}. In the case
of half-filling, with its conduction-electron band densely occupied
by one electron per local moment, the chemical potential falls in
between the hybridized bands, giving rise to a Kondo insulator (KI)
phase \cite{Aeppli-Fisk92,Ueda97,Riseborough00,Si-Paschen13}. The
KIs serve as a promising setting to realize correlated topological
insulator (TI) states \cite{Dzero10,Wolgast13,Kim13}. In addition,
they promise to yield novel types of quantum phase transitions
\cite{Yamamoto10,Feng13,Si-Paschen13}.

A regime which has hardly been explored theoretically is the dilute
carrier limit (DCL), where the conduction electron band is nearly
empty (or, equivalently, full). This is becoming a pressing issue,
as recent experimental studies have identified dilute-carrier heavy
fermion materials\cite{Luo15,Nakatsuji15}. In this work, we study
such a regime in an Anderson lattice model. We find a surprising KI
phase in the limit of empty conduction electron band. The emergence
of this phase crucially depends on the existence of a spin-orbit
coupling (SOC). It also is anchored by another phase that we
identify in the DCL, a Dirac-Kondo semimetal (DKSM) whose
quasiparticle excitations exhibit a non-trivial Berry phase. The
latter also makes heavy fermion systems in the DCL as a means to
realize correlated form of Dirac semimetals
\cite{Wallace47,Semenoff84,Neto09}.

\section{Topological electronic states and Kondo effect}

A simple platform for topological electronic states is the honeycomb
lattice as in graphene \cite{Geim07}. In the absence of the SOC, the
system at half filling or with one electron per site realizes a
Dirac semimetal with the conduction and valence bands touching at a
point in momentum space, the Dirac point, where the quasiparticles
show linear dispersions of Dirac
fermions\cite{Wallace47,Semenoff84,Neto09}. This semimetallic phase
can be topologically non-trivial if the corresponding quasiparticle
wavefunction carries a destructive $\pi$-Berry
phase\cite{Mikitik99,Xiao10}. The topological effect gives rise to
unconventional quantum Hall plateaus when a magnetic field is
applied as manifested in graphene\cite{Novoselov05,Zhang05}.
Remarkably, such topological semimetal can be viewed as a parent
phase of topological band insulators, in the sense that it can open
a bulk band gap and exhibit gapless surface states in the presence
of intrinsic SOC\cite{Kane-Mele05,Hasan-Kane10,Qi-Zhang11}. An
important open issue is what happens to such topological
semimetallic or insulating phases when the electron correlations are
strong\cite{Meng10,Rachel-Hur10,HLA11,YXL11,ZZW11,Kotov12}.

Meanwhile, strong correlations are known to play a dominant role in
KIs \cite{Hewson93,Coleman07,Si-Paschen13}. These systems are
emerging as the playing ground to study quantum criticality and
emergent phases due to the interplay between the Kondo effect and
Ruderman-Kittel-Kasuya-Yosida (RKKY) interaction
\cite{Yamamoto10,Pixley15}. The SOC yields non-trivial topology in
the electronic states, and allows for KIs that are topologically
nontrivial \cite{Dzero10,Dzero12,Lu13,Weng14}. A prime candidate for
such a topological Kondo insulator (TKI) is the paramagnetic
compound,
SmB$_6$\cite{Wolgast13,Kim13,Zhang13,Li13,Kim14,Chen13,Neupane13,Jiang13,Xu14}.
Prima facie one expects that the TKIs could emerge out of
topological band insulators just by turning on the Kondo coupling.
However, this is prevented by the topological no-go theorem, i.e.,
any two distinct topological non-trivial insulating phases cannot be
smoothly connected by a single tunable
parameter\cite{Hasan-Kane10,Qi-Zhang11}.

Motivated by the recent experiments on the dilute-carrier heavy
fermion materials, as well as by an earlier study about the role of
SOC in Kondo lattice systems\cite{Feng13}, in this paper we explore
the interplay between the Kondo effect and SOC in the DCL regime. We
investigate the paramagnetic phases in a honeycomb Anderson lattice
model at arbitrary electron fillings. We study the model when the
$f$-electron Coulomb repulsion is strong, and compare the results
with their counterparts when the interaction is weak. We demonstrate
the existence of the paramagnetic semimetallic phase, the DKSM, in
the absence of the SOC. When the SOC becomes nonzero, this phase
becomes the $Z_2$-TKI or -topological mixed valence insulator. We
also discuss the role of topological Berry phase and the Dirac
fermion behavior of low energy excitations.

\section{Anderson Lattice Model}
We start from the Hamiltonian for a honeycomb Anderson lattice:
\begin{eqnarray}\label{Anderson}
H=H_{0}+H_{cd}+H_d.
\end{eqnarray}
Here,
$H_0=-t\sum_{<\textbf{i},\textbf{j}>}(c^{\dagger}_{\textbf{i},\sigma}
c_{\textbf{j},\sigma}+h.c.)$ is the kinetic term of the conduction
$c$ electrons with the nearest-neighbor (n.n.) hopping,
$H_d=E_0\sum_{\textbf{i}\sigma}n_{d\textbf{i}\sigma}+
U\sum_{\textbf{i}}n_{d\textbf{i}\uparrow}n_{d\textbf{i}\downarrow}$
describes the energy level $E_0$ and  on-site Coulomb repulsion $U$
of the local $d$ electrons, and $H_{cd}=V\sum_{\textbf{i}\sigma}
(c^{\dag}_{\textbf{i}\sigma}d_{\textbf{i}\sigma}+h.c.)$ is the
hybridization between the $c$ and $d$ electrons. We will consider
$E_0$ to be well below the Fermi energy. Our primary interest is for
large $U$, where the model is mapped to a Kondo lattice. To gain
insights into the topological characteristics of the electronic
states, we will first consider the small $U$ regime.

In the momentum ${\bf k}$-space, the conduction electron part
shifted by the chemical potential $\mu$ takes the form
$H_{0}=\sum_{\textbf{k}\sigma}C_{\textbf{k}\sigma}^{\dag}M_{\textbf{k}}C_{\textbf{k}\sigma}$,
 with $C_{\textbf{k}\sigma}^{\dag} =
(c_{a,\textbf{k}\sigma}^{\dag},c_{b,\textbf{k}\sigma}^{\dag})$ and
\begin{eqnarray}
M_{\textbf{k}}=\left(\begin{array}{cc}
                                             -\mu & -g_{\textbf{k}} \\
                                           -g_{\textbf{k}}^* & -\mu \\
                                         \end{array}
                      \right),
\end{eqnarray}
where,
$g_{\textbf{k}}=t[1+e^{-i\textbf{k}\cdot\textbf{a}_1}+e^{-i\textbf{k}\cdot\textbf{a}_2}]$,
with $\textbf{k}$ valued in the first Brillouin zone (BZ). The
subscripts $a$ and $b$ denote two sublattices of the honeycomb
lattice where the primitive vectors are represented by
$\textbf{a}_1$ and $\textbf{a}_2$, respectively.

\section{Semimetal phase: the case of weak coupling}

We first consider the weak-coupling regime where the effect of $U$
can be treated perturbatively. We assume  that the $d$-orbital (with
$E_0<E_F$ ) is half-filled in the absence of the $c$-$d$
hybridization.  While a finite $V$ could delocalize the $d$
electrons, a small $U$ will lead to a renormalization of
quasi-particle weight $Z=\frac{1}{1-\frac{\partial
\Sigma_d(\omega)}{\partial \omega}|_{\omega=0}}$ for $d$-electrons.
Here, the self-energy of $d$-electrons is assumed to be
momentum-independent, $\Sigma_d(\vec {k}, \omega)\simeq
\Sigma_d(\omega)$, since the $d$-electrons are predominately
localized in the absence of the $c$-$d$ hybridization. The value of
$0\leq Z\leq 1$, which could be calculated perturbatively, leads to
an effective hybridization ${\tilde V}={\sqrt Z}V$. A crucial
feature of the honeycomb lattice is that, at the half-filling,
$\tilde{V}$ is non-zero only when $V$ is larger than a threshold
value, $V_c$, due to the Dirac-like dispersion of the conduction
band\cite{Feng13}. Usually, away from half-filling, $V_c$ is
significantly reduced due to an enhanced density of states near the
Fermi energy. The situation with a finite effective hybridization or
Kondo effect corresponds to the regime where $V>V_c$.

In this hybridization phase, the dressed ${\tilde
d}_{\textbf{k}\sigma}=\sqrt{Z}d_{\textbf{k}\sigma}$ is used for both
sublattices. Then, the Hamiltonian can be re-expressed by $H
=\sum_{\textbf{k}\sigma}\Psi_{\textbf{k}\sigma}^{\dag}H^{(weak)}_{\textbf{k}}\Psi_{\textbf{k}\sigma}$,
with
\begin{eqnarray}\label{matrix}
H^{(weak)}_{\textbf{k}}=\left(
                       \begin{array}{cc}
                         M_{\textbf{k}} & \tilde{V}\cdot I_{2\times 2} \\
                         \tilde {V}\cdot I_{2\times 2} & \epsilon_d \cdot I_{2\times 2} \\
                       \end{array}
                     \right).
\end{eqnarray}
Here,
$\Psi^{\dagger}_{\textbf{k}\sigma}=(c_{a,\textbf{k}\sigma}^{\dag},
c_{b,\textbf{k}\sigma}^{\dag},{\tilde
d}_{a,\textbf{k}\sigma}^{\dag},{\tilde
d}_{b,\textbf{k}\sigma}^{\dag})$, $I_{2\times 2}$ is a $2\times 2$
identity matrix, $\epsilon_d=Z(E_{0}+\Sigma'_d(0))$ is the
renormalized $d$ electron level. The eigenstates of the above
Hamiltonian are degenerate for the two spin components. Each of them
has four quasiparticle bands with energies
\begin{eqnarray}
E^{(\eta,\tau)}_{\textbf{k}}=
\frac{1}{2}\left(\epsilon_d+\epsilon^{(\eta)}
_{\textbf{k}}+\tau\sqrt{[\epsilon_d-\epsilon^{(\eta)}_{\textbf{k}}]^{2}+4{\tilde
V}^2}\right) \label{energy}.
\end{eqnarray}
Where, $\epsilon^{(\eta)}_{\textbf{k}}=\eta |g_{\textbf{k}}|-\mu$,
$\eta=(+, -)$ indicates the band index due to the superposition of
the (a,b) sublattices, $\tau=(+, -)$ the band index due to the
$c$-$d$ hybridization.

The order of the bands is related to the sign of
$\epsilon_d-\epsilon^{(\eta)}_{\textbf{k}}$. For convince, we denote
$E^{(1)}_{\textbf{k}}\equiv E^{(+,+)}_{\textbf{k}}$,
$E^{(2)}_{\textbf{k}}\equiv E^{(-,+)}_{\textbf{k}}$,
$E^{(3)}_{\textbf{k}}\equiv E^{(+,-)}_{\textbf{k}}$, and
$E^{(4)}_{\textbf{k}}\equiv E^{(-,-)}_{\textbf{k}}$. Without losing
generality, we assume the order $E^{(1)}_{\textbf{k}}\geq
E^{(2)}_{\textbf{k}}> E^{(3)}_{\textbf{k}}\geq E^{(4)}_{\textbf{k}}$
in the absence of hybridization. This is satisfied if the bare
$d$-electron level $E_0$ is at or below the bottom of the conduction
band, because for $\tilde{V}=0$ and
$\epsilon_d<\epsilon^{(\eta)}_{\textbf{k}}$ we have
$E^{(1)}_{\textbf{k}}=\epsilon^{(+)}_{\textbf{k}}$,
$E^{(2)}_{\textbf{k}}=\epsilon^{(-)}_{\textbf{k}}$,
$E^{(3)}_{\textbf{k}}=E^{(4)}_{\textbf{k}}=\epsilon_{d}$.

At the half-filling, $\mu=0$, $\epsilon^{(+)}_{\textbf{k}}$ and
$\epsilon^{(-)}_{\textbf{k}}$ contact at the Dirac points,
$\pm\textbf{k}_D=\pm\frac{2\pi}{3a_0}(1,\frac{1}{\sqrt 3})$ ($a_0$
is the distance of the two n.n. sites), so that
$g_{\pm\textbf{k}_D}=0$. In the hybridization phase $\tilde{V}>0$,
the four bands deform differently. An important observation is that
the order of bands $E^{(1)}_{\textbf{k}}\geq E^{(2)}_{\textbf{k}}>
E^{(3)}_{\textbf{k}}\geq E^{(4)}_{\textbf{k}}$ holds for arbitrary
$\tilde{V}>0$ and $\textbf{k}\in$ BZ. The equality holds only at the
Dirac points $\pm \textbf{k}_{D}$\cite{note}. Hence the two upper
bands $E^{(1)}_{\textbf{k}}$ and $E^{(2)}_{\textbf{k}}$, as well as
the two lower bands $E^{(3)}_{\textbf{k}}$ and
$E^{(4)}_{\textbf{k}}$, contact at the Dirac points respectively.
Such band touching behavior is robust for any $\tilde V>0$ since the
hybridization keeps the required discrete symmetries like those in
graphene. However, the two intermediate bands $E^{(2)}_{\textbf{k}}$
and $E^{(3)}_{\textbf{k}}$ are always separated by a finite gap
$\Delta=\sqrt{(\epsilon_d+\mu)^2+4{\tilde V}^2}$. This is the
hybridization gap which leads to the Kondo or mixed valence
insulating phase at the half filling ($\mu=0$).

If we tune the chemical potential upward to $\mu=\mu_c=
\frac{\tilde{V}^2}{|\epsilon_d|}$, the Fermi energy is shifted to
the Dirac points so that
$E^{(1)}_{\pm\textbf{k}_D}=E^{(2)}_{\pm\textbf{k}_D}=0$. The total
electron filling factor is then $\frac{n}{2}=(n_c+n_d)/2=3/4$.
Notice that the average particle number per unit cell $n_c$ (or
$n_d$) alone is not conserved when $\tilde{V}\neq 0$, and can be
calculated by $n_c=-\frac{1}{2{\cal N}}\frac{\partial}{\partial
\mu}\sum_{\textbf{k},i=2,3,4} E^{(i)}_{{\textbf k}}$ with ${\cal N}$
being the total number of unit cells. When $\tilde{V}=0$, the total
electron filling $n/2=3/4$ corresponds to the case with $n_c=2$ (the
full-filling for $c$ electrons) and $n_d=1$ (the half-filling for
$d$ electrons).

Figure 1(a) shows the band structure along some high symmetry points
($\Gamma\rightarrow M\rightarrow K\rightarrow \Gamma$) in the
weak-coupling regime where the $K$-point is one of the Dirac points
$\pm\textbf{k}_{D}$. Here the parameters $t=1$ and $E_0=-3$ are
fixed. Without loss of generality, we also choose $\epsilon_d=-2.5,
\tilde V=0.9$ as due to the renormalization effect of $U$. In
Fig.1(a) we shift the Fermi energy at the Dirac points in between
the two upper bands ( where $E_F=0$). This is realized by increasing
the chemical potential to $\mu_c=0.324$. The hybridization gap opens
in between the second and third bands. The flat band nature of the
$d$-level is mainly reflected in the two lower bands.
\begin{figure}
\includegraphics[width=9.0cm]{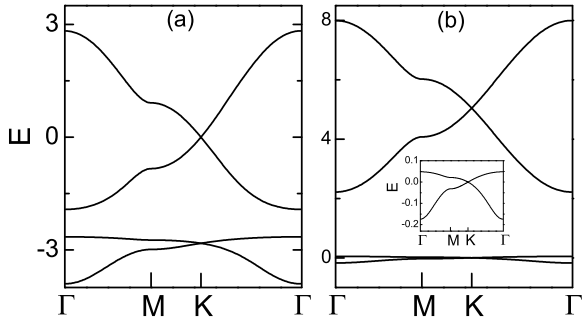}
\caption{Band structure along the high symmetry points. (a) Case
(I): the weak-coupling regime at the $3/4$-filling. (b) Case (II):
the strong-coupling regime at the $1/4$-filling. The inset in (b)
shows the two lower bands in an enlarged scale. In both cases,
$t=1$, $E_0=-3$, but the hybridization $V$ and chemical potential
$\mu$ are different as described in the text. The Fermi energy is
shifted to the Dirac points by tuning the chemical potential $\mu>0
(<0)$ in Case (I) and Case (II) accordingly. }
\end{figure}

In principle the Fermi energy can be also shifted downward by tuning
the chemical potential. It would be expected that we can tune the
total electron filling factor to
$\frac{n}{2}=(\frac{0}{2}+\frac{1}{2})/2=1/4$ so that the Fermi
energy is shifted to $E^{(3)}_{\pm\textbf{k}_{\textbf
D}}=E^{(4)}_{\pm \textbf{k}_{\textbf D}}$. But this value is always
negative as $\epsilon_d <0$ and $\mu>\epsilon_d$ in the
weak-coupling regime. We will see below that the Fermi energy can be
tuned to zero exactly in the strong-coupling regime.

\section{Semimetal phase: the case of strong coupling}

We turn next to the strong-coupling regime where $U$ is large enough
so that no double occupancy is allowed
 on each sites of the $d$-sublattice. We utilize the slave-boson method
\cite{Hewson93}, and express
$d^{\dag}_{\textbf{i}\sigma}=f^{\dag}_{\textbf{i}\sigma}b_{\textbf{i}}$,
with $f^{\dag}_{\textbf{i}\sigma}$ and $b_{\textbf{i}}$ being
respectively fermionic and bosonic operators satisfying the
constraint $b^{\dag}_{\textbf{i}}b_{\textbf{i}}+\sum_{\sigma}
f^{\dag}_{\textbf{i}\sigma}f_{\textbf{i}\sigma}=1$. Introducing the
basis
$\Psi^{\dagger}_{\textbf{k}\sigma}=(c_{a,\textbf{k}\sigma}^{\dag},
c_{b,\textbf{k}\sigma}^{\dag},f_{a,\textbf{k}\sigma}^{\dag},f_{b,\textbf{k}\sigma}^{\dag})$
, the  mean-field Hamiltonian is expressed as $H_{MF}
=\sum_{\textbf{k}\sigma}\Psi_{\textbf{k}\sigma}^{\dag}H^{(strong)}_{\textbf{k}}\Psi_{\textbf{k}\sigma}$
with
\begin{eqnarray}\label{Matrix}
H^{(strong)}_{\textbf{k}}=\left(
                       \begin{array}{cc}
                         M_{\textbf{k}} & rV\cdot I_{2\times 2} \\
                         rV\cdot I_{2\times 2} & (E_0+\xi)\cdot I_{2\times 2} \\
                       \end{array}
                     \right).
\end{eqnarray}
Here, $r=\langle b\rangle$ is the condensation density of the
bosons, and $\xi$ is the Lagrange multiplier imposing the
constraint.

The structure of the strong-coupling Hamiltonian Eq.(\ref{Matrix})
is similar to  that in the weak-coupling regime, Eq.(\ref{matrix}).
The quasiparticle bands are then obtained as in Eq.(\ref{energy})
with replacements $\epsilon_d\rightarrow E_0+\xi$ and $\sqrt
Z\rightarrow r$. Hence in the hybridization phase $V^*=rV\neq 0$, or
$r>0$, there are four quasiparticle bands, with two upper bands and
two lower bands contact respectively at the Dirac points. Different
from the weak-coupling regime, the parameters $r$ and $\xi$ should
be determined self-consistently by $\frac{1}{2{\cal
N}}\sum_{\textbf{k}\sigma;\alpha=a,b}\langle
f^{\dag}_{\alpha,\textbf{k}\sigma}f_{\alpha,\textbf{k}\sigma}\rangle+
r^2 =1$, $\frac{V}{2{\cal
N}}\sum_{\textbf{k}\sigma;\alpha=a,b}\Re{\langle
c^{\dag}_{\alpha,\textbf{k}\sigma}f_{\alpha,\textbf{k}\sigma}\rangle}+r\xi=0
$, with $\Re$ indicating the real part. Here we numerically solve
these equations for $t=1, E_0=-3$ for various $V$ at the fractional
fillings of the total electron $n/2 =(n_c+n_d)/2$ per unit cell,
with $n_c=-\frac{1}{2\cal N}\frac{\partial}{\partial
\mu}\sum_{\textbf{k},i} E^{(i)}_{{\textbf k}}$ and
$n_d=\frac{1}{2\cal N}\frac{\partial}{\partial
\xi}\sum_{\textbf{k},i} E^{(i)}_{{\textbf k}}$ for occupied band(s)
$i$.

Similar to the half-filling case where a finite value of $V_c$ is
required for the Kondo phase, we find that at the 1/4-filling the
solution for $r>0$ exists when $V>V_c\sim 2.6$. To shift the Fermi
energy to zero at the Dirac points, we can tune the chemical
potential to $\mu_c=-\frac{r^2V^2}{E_0+\xi}$ so that $|\mu|$ is
relatively large in the strong-coupling regime. In this regime,
$E_0+\xi$ is slightly above the Fermi energy so that the Kondo phase
is stabilized. For example, a set of self-consistent parameters
$\xi=3.129$, $r=0.241$, and $\mu= - 4.903$ is obtained when we fix
$V=3.3$.   The resulting band structure is shown in Fig.1(b) where
the four bands for high symmetry points are plotted. Although the
two lower bands are relatively narrow, they contact at the Dirac
points where the Fermi energy locates realizing the DKSM phase. In
principle, the Fermi energy can be also shifted upward. But
precisely at the $3/4$-filling, only the trivial solution $r=0$
exists, indicating that the ground state is always in the decoupled
phase. This is due to the strong on-site $U$.

Having established its existence in the weak- and strong-coupling
regime, we expect that the semimetal phase also exist for
intermediate values of $U$. On general grounds, the
symmetry-preserving nature of the hybridization we consider ensures
that the location of the Dirac points is protected by the inversion
symmetry of the lattice, as in graphene. More microscopically, in
the Kondo-hybridized phase, the strength of hybridization is
expected to interpolate between the results calculated in the small
and large $U$ limits. In other words, within the hybridized phase,
the main effect of varying $U$ is expected to renormalize the model
parameters.

\section{Topological Berry phase and Dirac heavy fermions}

Our analysis so far establishes the existence of the paramagnetic
semimetal phase in the present honeycomb Anderson lattice away from
half-filling. The extreme situations include Case (I): in the
weak-coupling regime at the $3/4$-filling, and Case (II): in the
strong-coupling regime at the $1/4$-filling. In these two cases the
Fermi energy can be tuned exactly to zero. The next issue is whether
the semimetallic phase is topologically non-trivial. We thus
calculate the Berry phase of the quasiparticle wavefunctions around
the Dirac points. For instance, let $|\Phi_{\textbf{k}}\rangle=\{
\Phi_{\textbf{k}}^{(1)},\cdots, \Phi_{\textbf{k}}^{(4)}\}^{T}$ be
the eigenstate of $h_{\textbf{k}}$, with $\Phi_{\textbf{k}}^{(i)}$
given explicitly in the SM. When the crystal momentum $\textbf{k}$
rotates adiabatically with a time-like parameter $\lambda$
 in a loop ${\cal C}$,
$|\Phi_{\textbf{k}}\rangle$ is always single-valued, while the
corresponding Berry phase is given by
\begin{eqnarray}
\theta=-i\oint_{\cal
C}\langle\Phi_{\textbf{k}}|\frac{\partial}{\partial
\lambda}|\Phi_{\textbf{k}}\rangle d\lambda.
\end{eqnarray}
We find that when $\cal C$ encloses the Dirac point $\textbf{k}_D$,
$\theta=2\pi\cdot[|\Phi_{\textbf{k}}^{(1)}|^2+|\Phi_{\textbf{k}}^{(3)}|^2]
=\pi$. Similarly, $\theta=-\pi$ when $\cal C$ encloses the Dirac
point $-\textbf{k}_D$. Notice that the above derivation holds for
both weak- and strong-coupling regimes. Such a non-trivial Berry
phase will cancel the contribution from the zero-point quantum
fluctuation\cite{Mikitik99,Xiao10}, leading to the shifted quantum
Hall plateaus as observed in graphene\cite{Novoselov05,Zhang05}.

It is also interesting to ask whether the low energy excitations
near the Dirac points are Dirac fermions. In the Kondo-destroyed
phase, the semimetal appears at half-filling like in graphene so
that the low energy excitations are Dirac fermions, exhibiting the
relativistic dispersion $\epsilon^{(\eta)}_{\textbf{q}}=\eta
v_F|\textbf{q}|$ for small $|\textbf{q}|$ shifted from
$\pm\textbf{k}_{D}$, with the velocity $v_F=\frac{3t}{2}a_0$. In the
Kondo hybridized phase, the corresponding low-energy excitations
show different behavior. In general, quadratic dispersions emerge
due to the finite effective hybridization. For illustration, we
consider the situation where $V$ is close to the phase boundary in
Case (I) or Case (II) where $\tilde V$ or $V^*$ is small. The
particle- and hole-like excitations  take the form
$E^{(\pm)}_{\textbf{q}}=\pm {\tilde v}_F|\textbf{q}|+{\tilde
a}\textbf{q}^2$ up to ${\cal O}
(|\frac{\textbf{q}}{\textbf{k}_{D}}|^2)$ in Case (I), and
$E^{(\pm)}_{\textbf{q}}=\pm v^*_F|\textbf{q}|-a^*\textbf{q}^2$ in
Case (II), with the corresponding renormalized Fermi velocities
${\tilde v}_F$, $v^*_F$ and the momentum-dependent coefficients
${\tilde a}$, $a^*$ given explicitly in the Supporting Information
\cite{note}.

In the immediate vicinity of the momentum space near the Dirac
points, the linear term dominates so that the low energy excitations
are approximately Dirac fermions. For Case (I), when hybridization
increases, the renormalized Fermi velocity ${\tilde v}_F$ decreases
and the quadratic term ${\tilde a}$  increases. Hence, the range of
the shifted momentum $\textbf{q}$ where the linear dispersion
dominates is reduced by the effective hybridization. This implies
that the excitations gain a finite effective mass beyond this range.
For Case (II), the suppression of the renormalized Fermi velocity
$v^{*}_F$ is stronger so that $a^*$ is effectively enhanced.  As a
result, the quasiparticle band $E^{(3)}_{\textbf{k}}$ ( or quasihole
band $E^{(4)}_{\textbf{k}}$ ) is significantly flattened due to the
$f$-electron feature, indicating that the momentum range where the
energy dispersion exhibits linear behavior is narrowed. Therefore
the effective mass is enhanced in the DKSM phase. For this reason,
we can also call the quasiparticles in the DKSM as Dirac heavy
fermions (DHFs).
\begin{figure}
\includegraphics[width=7.0cm]{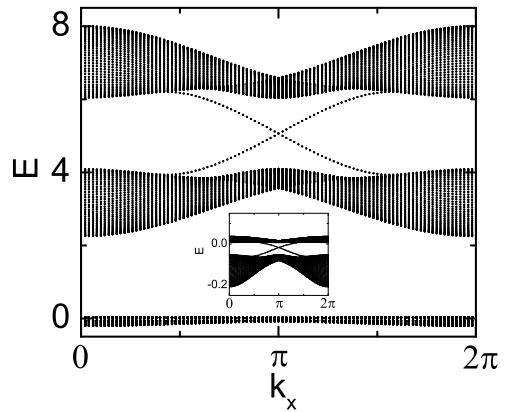}
\caption{Slave-boson spectra of a strip of honeycomb lattice along
$x$-direction for Case (II). The model parameters are $E_0=-3$,
$t=1$, $V=3.3$, and $\lambda_{so}=0.3$. Other self-consistently
solved parameters are $\xi=3.128$, $r=0.256$, and $\mu= - 4.913$.
The inset is for the two lower bands with enlarged scale. }
\label{edge}
\end{figure}

\section{Topological Kondo insulator}

We now discuss the influence of the conduction electrons' SOC on the
semimetal phase. We add $H_{SO}= i\lambda_{so}\sum_{\ll
\textbf{i}\textbf{j}\gg\sigma\sigma'}v_{\textbf{i}\textbf{j}}c^{\dagger}_{\textbf{i}\sigma}
{\sigma}^z_{\sigma\sigma'} c_{\textbf{j}\sigma'}$ to
Eq.(\ref{Anderson}), with $\lambda_{so}$ being the strength of the
intrinsic (Dresselhaus-type) SOC of the conduction electrons and
$v_{\textbf{i}\textbf{j}}=\pm 1$ depends on the direction of hopping
between the next-nearest-neighbor sites\cite{Kane-Mele05,Feng13}. In
general, $\lambda_{so}$ will open a gap in the semimetal phase
\cite{note} and produce edge states. For example, in Case (II), we
show in Fig. 2 the energy spectra of a stripe of the finite system
with a zigzag geometry by using the same set of model parameters as
discussed previously. A finite band gap in the two lower (or two
upper) bands opens immediately with metallic edge states.

The above calculation demonstrates that the DKSM or hybridization
semimetal can be driven to TKI or topological mixed valence
insulator by turning on the SOC, as schematically shown in Fig. 3.
As demonstrated previously\cite{Feng13}, there is a phase transition
from the TI to KI phases when going across the dotted line on the
$\mu=0$ (or $n_c=1$)-plane corresponding to the half-filling case.
However, no direct transition can take place from the TI to TKI
phases by tuning only one of the three parameters. Of course, the
TKI phase could also be realized from the topological trivial KI
phase in the presence of the SOC by tuning the chemical potential,
or from the helical metal phase by tuning both the chemical
potential and Kondo coupling.
\begin{figure}
\includegraphics[width=6.0cm]{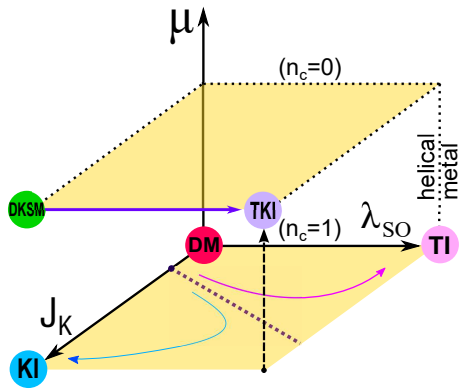}
\caption{(Color on line) Cartoon picture of phases in terms of the
spin-orbit coupling $\lambda_{so}$, Kondo coupling $J_K$, and the
chemical potential $\mu$. The Dirac metal (DM), topological
insulator (TI), and Kondo insulator (KI) phases are in the
$\mu=0$-plane (with $n_c$=1 or at half-filling). The Dirac-Kondo
semimetal (DKSM) and topological Kondo insulator (TKI) phases are in
the $\mu=1/4$(or $\mu=3/4$)-plane ( with $n_c=0$ or $2$
corresponding empty or full filling of conduction band). }
\label{phase}
\end{figure}

\section{New physics in the dilute limit: discussions}

Fig. 4 illustrates the DCL regime we are considering, with the local
moments represented by a Mott-insulator component and the conduction
electrons given by a band-insulator component. Our results highlight
several important features in this DCL regime of heavy fermion
systems. First, a relatively large $V_c$ as shown for Case (II) is
required in the Kondo phase, indicating the delayed Kondo screening
as a natural consequence of the so-called Nozi\`{e}res exhaustion
problem\cite{Nozieres98} in this limit. This conclusion is
consistent with the recent experiments, which show the depletion of
the coherent Kondo energy scale \cite{Luo15,Chen16}. Second, the
band touching and topological features are both inherited from those
of the itinerant and local components of the lattice model due to
the symmetry preserving hybridization. The band curvature deforms
drastically due to the finite hybridization, rendering the linear
Dirac dispersion valid only in a narrowed momentum region in the
vicinity of the Dirac points. Third, in the Kondo phase, the SOC of
conduction electrons can generate a bulk gap at the Dirac points
leading to a correlated TI state. Fourth, the surface states of the
TKI phase could be either light or heavy fermions, depending on the
details of the hybridization and the quasiparticle momentum. In this
connection, we notice that both light and heavy quasiparticles as
well topological non-trivial or trivial surface states are observed
in SmB$_6$ and YbB$_6$ depending on surface directions in several
recent magneto-thermoelectric transport and pressure
measurements\cite{Luo14,Zhou15,Yong15,Alexandrov15}. Finally, while
we have emphasized the SOC of the conduction electrons, an SOC of
$f$ electrons will also open a bulk gap driving the semimetal phase
to the TKI phase.

So far we considered the paramagnetic state, which has a prevailing
interest in understanding the fundamental underlying principle of
the electronic structures and can be stabilized by both SOC and
quantum fluctuations. In the absence of SOC, on the other hand, the
system could be ferromagnetically ordered in the DCL regime
\cite{Ueda97}. We can qualitatively explore the influence of the
ferromagnetic order by $ H_{J}=\sum_{{\bf k}\sigma,\alpha=a,b}\sigma
M[J_K c^{\dagger}_{\alpha,{\bf k}\sigma}c_{\alpha,{\bf k}\sigma}-I_R
f^{\dagger}_{\alpha,{\bf k}\sigma}f_{\alpha,{\bf k}\sigma}]$, with
parameters $J_K$, $I_R$ and $M$ being the effective Kondo coupling,
the RKKY interaction, and the magnetization, respectively. The
non-zero $M$ splits the spin degeneracy, leading to the shifts
$\mu\rightarrow \mu_{\sigma}=\mu-\sigma M J_K$ and $E_0\rightarrow
E_{0,\sigma}=E_0-\sigma M I_R$. Therefore, in each spin channel, the
order of four bands does not alter and the band touching feature at
the Dirac points $\textbf{k}_D$ is still robust. This feature
ensures the existence of the ferromagnetic Dirac semimetal phase by
properly tuning the spin-dependent chemical potentials. The relative
band fillings of each spin channel are delicately dependent on the
effective parameters $J_K$, $I_R$, and $M$. The consequence of these
deserves future investigations.

We close this section by discussing the experimental realizations of
our results. Given the recent emergence of heavy fermion systems in
the DCL regime, it is quite plausible to realize the novel type of
Kondo insulator states in heavy fermion compounds whose
conduction-electron band is insulating, especially those with the
non-$f$ electrons being hosted by heavy elements\cite{Rai}. In
addition, Fig.~4 also inspires us to propose a new type of structure
to realize the physics we have discussed. This corresponds to
building a heterostructure between a Mott insulator and a band
insulator. Such structures can be engineered so that i) for the band
insulator, the chemical potential is close to the top of the valence
band or the bottom of the conduction band and ii) the barrier at the
interface will be relatively small so that a nonzero hybridization
will operate between the electronic states illustrated in Fig.~4.
\begin{figure}
\includegraphics[width=8.0cm]{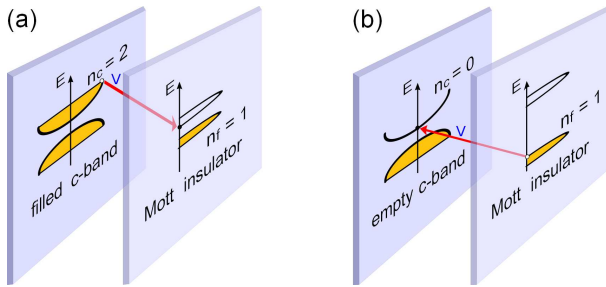}
\caption{(Color online) DKSM and KI phases emerge from the interface
of two different insulators due to a proximity effect. One is a Mott
insulator, which provides a lattice of local moments. The other is a
band insulator, with the chemical potential that is close to either
(a) top of the valence band or (b) bottom of the conduction band.}
\label{expsetup}
\end{figure}

\section{Summary and outlook}

To summarize, we have shown that the correlated semimetal phases
exist in the honeycomb Anderson lattice in the dilute limit of
charge carriers where the conduction electron band is nearly empty
or full. They could be realized in heavy fermion compounds in the
dilute-carrier limit, or at the interface between trivial band and
Mott insulators due to a proximity effect. The low energy
excitations in such semimetals are approximately linear in the
vicinity of the Dirac points. For the case with a large
$f$-electron's correlation and almost empty band of conduction
electrons, the DKSM is realized and the corresponding quasiparticles
are the DHFs because the $f$-electron's contribution is
inextricable. In particular, when the quasiparticle's momenta are
moderately away from the Dirac points, such DHFs will gain a finite
(cyclotron) mass due to the quadratic correction to the energy
dispersion, and manifest the heavy-fermion behavior.

We have also demonstrated that when turning on the SOC, the
hybridization semimetal or DKSM phase will open a bulk gap while
with metallic surface states. This phase is different to either KI
or TI phases at the half-filling case\cite{Feng13}. Thus, the
present study has established an alternative route towards the
formation of TKIs. In this scenario, a TKI is driven from the DKSM
phase by the SOC of conduction electrons. Our findings underscore
that the interplay between the Kondo coupling and SOC can in general
nucleate new quantum phases and their transitions in strongly
correlated settings.

\vskip 1cm

\begin{acknowledgments}

We would like to thank P. Goswami, Y. Luo, E. Morosan, S. Paschen,
and B. K. Rai for useful discussions. This work was supported in
part by the NSF of China (under Grants No. 11304071 and No.
11474082). Q.S. was supported in part by the ARO Grant No.
W911NF-14-1-0525 and the Robert A. Welch Foundation Grant No.
C-1411.

\end{acknowledgments}


\newpage
\setcounter{equation}{0}

\begin{center}
{\bf \Large SUPPORTING INFORMATION} \vskip 1.0 cm

\section*{Dirac-Kondo semimetals and
topological Kondo insulators in the dilute carrier limit}
\end{center}

\vskip 1.0 cm

\begin{center}
Xiao-Yong Feng$^{1}$, Hanting Zhong$^{1}$, Jianhui Dai$^{1}$, and Qimiao Si$^{2}$ \\
$^1${\it Condensed Matter Group, Department of Physics, Hangzhou
Normal University, Hangzhou 310036,
China}\\
$^2${\it Department of Physics and Astronomy, Rice University,
Houston, TX 77005, USA }
\end{center}

\vskip 1.0 cm
\renewcommand{\theequation}{S\arabic{equation}}

In this Supporting Information, we provide additional derivations
for the details of the band dispersion, demonstrate the robust of
the Dirac points in the hybridization phase in the absence of the
SOC, and further discuss the formation of topological Kondo or mixed
valence insulating state in the presence of the SOC of conduction
electrons. The dispersion relations of the low-energy excitations
near the Dirac points in the semimetal phase are also presented
explicitly.

The Hamiltonian we study is the Honeycomb Anderson lattice with the
intrinsic SOC of conduction electrons, described by
\begin{eqnarray}\label{AndersonSOC}
H&=&-t\sum_{\langle\textbf{i}\textbf{j}\rangle\sigma}c^{\dagger}_{\textbf{i}\sigma}
c_{\textbf{j}\sigma}+V\sum_{\textbf{i}\sigma}
(c^{\dag}_{\textbf{i}\sigma}d_{\textbf{i}\sigma}+h.c.) \nonumber\\
&&+E_0\sum_{\textbf{i}\sigma}d^{\dag}_{\textbf{i}\sigma}d_{\textbf{i}\sigma}
+U\sum_{\textbf{i}}n_{d\textbf{i}\uparrow}n_{d\textbf{i}\downarrow}\nonumber\\
&&+ i\lambda_{so}\sum_{\ll
\textbf{i}\textbf{j}\gg\sigma\sigma'}v_{\textbf{i}\textbf{j}}c^{\dagger}_{\textbf{i}\sigma}
{\sigma}^z_{\sigma\sigma'} c_{\textbf{j}\sigma'}.
\end{eqnarray}
The first two lines correspond to the Hamiltonian
Eq.(\ref{Anderson}) given in the main text, the third line is the
SOC of conduction electrons.

In the weak-coupling regime, there are totally eight bands of
eigenstates, with the following eigen energies
\begin{eqnarray}
E^{(\eta,\tau,\sigma)}_{\textbf{k}}=
\frac{1}{2}\left(\epsilon_d+h^{(\eta)}
_{\textbf{k}}-\mu+\tau\sqrt{[\epsilon_d-h^{(\eta,\sigma)}_{\textbf{k}}+\mu]^{2}+4{\tilde
V}^2}\right).\nonumber\\ \label{energysoc}
\end{eqnarray}
Where, $h^{(\eta,\sigma)}_{\textbf{k}}=\eta
\sqrt{|g(\textbf{k})|^2+\Lambda^2_{\textbf{k}\sigma}}$,
$g_{\textbf{k}}=t[1+e^{-i\textbf{k}\cdot\textbf{a}_1}+e^{-i\textbf{k}\cdot\textbf{a}_2}]$,
$\Lambda_{\textbf{k}\sigma}=2\sigma\lambda_{so}\{\sin[\textbf{k}\cdot\textbf{a}_1]
-\sin[\textbf{k}\cdot\textbf{a}_2]-\sin[\textbf{k}\cdot(\textbf{a}_1-\textbf{a}_2)]\}$.
$\sigma=(+1,-1)$ indicates the spin index due to the spin up and
spin down components, $\eta=(+, -)$ indicates the band index due to
the superposition of the (a,b) sublattices, and $\tau=(+, -)$ the
band index due to the $c$-$d$ hybridization. The eigenstates with
$\sigma=(+1,-1)$ are degenerate due to the time reversal symmetry,
resulting in four energy bands  $E^{(i)}_{\textbf{k}}$ (
$i=1,\cdots,4$) in each spin sector, as given explicitly in Eq.(S3)
where the index $\sigma$ in Eq.(\ref{energysoc}) is dropped in the
following discussions. For a given band $E^{(i)}_{\textbf{k}}$, the
corresponding eigenstate can be solved as
$|\Phi_{\textbf{k}}\rangle=\{ \Phi_{\textbf{k}}^{(1)},\cdots,
\Phi_{\textbf{k}}^{(4)}\}^{T}$, where
$\Phi_{\textbf{k}}^{(1)}=\frac{1}{m_{\textbf{k}}}
\frac{\tilde{V}^2-(E^{(i)}_{\textbf{k}}+\mu)(E^{(i)}_{\textbf{k}}-\epsilon_d)}{\tilde{V}tg^{*}_{\textbf{k}}}$,
$\Phi_{\textbf{k}}^{(2)}=\frac{1}{m_{\textbf{k}}}
\frac{(E^{(i)}_{\textbf{k}}-\epsilon_d)}{\tilde{V}}$,
$\Phi_{\textbf{k}}^{(3)}=\frac{1}{m_{\textbf{k}}}
\frac{\tilde{V}^2-(E^{(i)}_{\textbf{k}}+\mu)(E^{(i)}_{\textbf{k}}-\epsilon_d)}{(E^{(i)}_{\textbf{k}}-\epsilon_d)tg^{*}_{\textbf{k}}}$,
and $\Phi_{\textbf{k}}^{(4)}=\frac{1}{m_{\textbf{k}}}$, with
$m_{\textbf{k}}$ being a normalization coefficient.

In the strong-coupling regime, the quasiparticle bands can be solved
in a similar manner following the weak-strong correspondence
illustrated in the main text. The situation $\mu=0$ in the
strong-coupling regime was considered in Ref.\cite{Feng13},
corresponding to the half-filling of both conduction and local
electrons. We emphasize here that in the strong-coupling regime, the
solutions obtained at $\mu=0$ do not necessary apply to the case of
$\mu\neq 0$ because the mean-field parameters such as the boson
density $r$ and the Lagrangin multiplier $\xi$, which must be solved
self-consistently, are strongly dependent on the chemical potential
$\mu$. On the other word, the rigid band shift of band structure,
which usually applies to the weak-coupling regime, does not applies
to the strong-coupling regime. Indeed, the band structures display
drastic deformation at the $1/4$- or $3/4$-filling and no solution
exists at the strong-coupling for the $1/4$-filling case where the
Fermi energy is precisely at the Dirac points. We here mainly focus
on the semimetal phase and the related quasiparticle properties
assuming the existence of the aforementioned mean-field parameters.

\section*{In the absence of spin-orbital coupling}

In the absence of the SOC, we denote
\begin{eqnarray}
E^{(1)}_{\textbf{k}}(\tilde V)&=& \frac{1}{2}\left(\epsilon_d+|g
_{\textbf{k}}|-\mu+\sqrt{[\epsilon_d-|g_{\textbf{k}}|+\mu]^{2}+4{\tilde
V}^2}\right),\nonumber\\
E^{(2)}_{\textbf{k}}(\tilde V)&=& \frac{1}{2}\left(\epsilon_d-|g
_{\textbf{k}}|-\mu+\sqrt{[\epsilon_d+|g_{\textbf{k}}|+\mu]^{2}+4{\tilde
V}^2}\right),\nonumber\\
E^{(3)}_{\textbf{k}}(\tilde V)&=& \frac{1}{2}\left(\epsilon_d+|g
_{\textbf{k}}|-\mu-\sqrt{[\epsilon_d-|g_{\textbf{k}}|+\mu]^{2}+4{\tilde
V}^2}\right),\nonumber\\
E^{(4)}_{\textbf{k}}(\tilde V)&=& \frac{1}{2}\left(\epsilon_d-|g
_{\textbf{k}}|-\mu-\sqrt{[\epsilon_d+|g_{\textbf{k}}|+\mu]^{2}+4{\tilde
V}^2}\right).\nonumber\\
\end{eqnarray}
Where, $E^{(1)}_{\textbf{k}}=E^{(+,+)}_{\textbf{k}}$,
$E^{(2)}_{\textbf{k}}=E^{(-,+)}_{\textbf{k}}$,
$E^{(3)}_{\textbf{k}}=E^{(+,-)}_{\textbf{k}}$,
$E^{(4)}_{\textbf{k}}=E^{(-,-)}_{\textbf{k}}$. Then, we have the
following propositions.

{\textbf{Proposition I}. If $E^{(1)}_{\textbf{k}}(0)\geq
E^{(2)}_{\textbf{k}}(0)\geq E^{(3)}_{\textbf{k}}(0) \geq
E^{(4)}_{\textbf{k}}(0)$ for any $\textbf{k}\in$ BZ, then
$E^{(1)}_{\textbf{k}}(\tilde V)\geq E^{(2)}_{\textbf{k}}(\tilde V)>
E^{(3)}_{\textbf{k}}(\tilde V) \geq E^{(4)}_{\textbf{k}}(\tilde V)$
for any finite $\tilde {V}$ and $\textbf{k}\in$ BZ.

{\textbf{Proposition II(a)}. In the weak-coupling regime and at
$3/4-$filling, when $\tilde V$ is sufficiently small, the particle-
and hole-like excitations take the form
\begin{eqnarray}
E^{(\pm)}_{\textbf{q}}=\pm {\tilde v}_F|\textbf{q}|+{\tilde
a}\textbf{q}^2
\end{eqnarray}
up to ${\cal O} (|\frac{\textbf{q}}{\textbf{k}_{D}}|^2)$, with
${\tilde v}_F=[1-\frac{{\tilde V}^2}{|\epsilon_d+\mu|^2}]v_F$,
${\tilde a}=[\frac{{\tilde
V}^2}{|\epsilon_d+\mu|^3}\mp(1-\frac{{\tilde
V}^2}{|\epsilon_d+\mu|^2})\frac{\sin(3\theta_{\textbf{q}})}{6t}]v^2_{F}$,
and $\theta_{\textbf{q}}=\arctan(\frac{q_x}{q_y})$.

{\textbf{Proposition II(b)}. In the strong-coupling regime and at
$1/4-$filling, when $V^*$ is sufficiently small, the particle- and
hole-like excitations take the form
\begin{eqnarray}
E^{(\pm)}_{\textbf{q}}=\pm v^*_F|\textbf{q}|-a^*\textbf{q}^2,
\end{eqnarray}
with $ v^*_F=\frac{V^{*2}}{|E_0+\xi+\mu|^2}v_F $,
${a^*}=[\frac{V^{*2}}{|E_0+\xi+\mu|^3}\pm
\frac{V^{*2}}{|E_0+\xi+\mu|^2}\frac{\sin(3\theta_{\textbf{q}})}{6t}]v^2_{F}$.

Proof. The effective hybridization $V$ keeps the discrete symmetries
such as the time reversal symmetry, the lattice inversion symmetry
as well the $C_{3v}$ rotation symmetry. Therefore, the existence of
the Dirac points $\textbf{k}_{D}$ is robust, where
\begin{eqnarray*}
E^{(1)}_{\textbf{k}_D}=E^{(2)}_{\textbf{k}_D}=\frac{1}{2}[\epsilon_d
-\mu+\sqrt{(\epsilon_d+\mu)^2+4 \tilde V ^2}],\\
E^{(3)}_{\textbf{k}_D}=E^{(4)}_{\textbf{k}_D}=\frac{1}{2}[\epsilon_d
-\mu-\sqrt{(\epsilon_d+\mu)^2+4 \tilde V ^2}].
\end{eqnarray*}
So at the Dirac points, the two pairs of bands
$E^{(1)}_{\textbf{k}}$ and $E^{(2)}_{\textbf{k}}$,
$E^{(3)}_{\textbf{k}}$  and $E^{(4)}_{\textbf{k}}$ contact to each
other. But the bands $E^{(2)}_{\textbf{k}}$ and
$E^{(3)}_{\textbf{k}}$ are separated at the Dirac points by a finite
gap $\Delta=\sqrt{(\epsilon_d+\mu)^2+4 \tilde V ^2}$.

It is further observed that when tuning the chemical potential $\mu$
so that the Fermi energy is shifted to $E^{(1)}_{\textbf{k}_D}$=
$E^{(2)}_{\textbf{k}_D}=0$, one must have $\mu>\epsilon_d$. This is
realized when $\mu_c=-\frac{\tilde V ^2}{\epsilon_d}$, corresponding
to the case of $3/4$-filling of conduction band. Similarly, the
Fermi energy can be shifted to $E^{(3)}_{\textbf{k}_D}$=
$E^{(4)}_{\textbf{k}_D}=0$ only when $\mu<\epsilon_d$ where the
conduction band filling is $1/4$. This can be realized in the
strong-coupling regime where $E_0+\xi>0$ and $E_0+\xi+\mu<0$. the
chemical potential is $\mu_c=-\frac{V^{*2}}{E_0+\xi}$.

Now we consider $\textbf{k}$ near the Dirac points so that
$|g_{\textbf{k}}/A|$ is sufficiently small, where $A=\epsilon_d+\mu$
(or $A=E_0+\xi+\mu$) in the weak-coupling regime (or the
strong-coupling regime).  Up to the order of ${\cal
O}(|g_{\textbf{k}}/A|^2)$, we have
\begin{eqnarray}
E^{(\lambda,\tau)}_{\textbf{k}}&=&\frac{A+\tau
\sqrt{A^2+4\tilde{V}^2}}{2}-\mu\\
&+&\frac{\lambda}{2}(1-\frac{\tau
A}{\sqrt{A^2+4\tilde{V}^2}})g_{\textbf{k}}+\frac{\tau
\tilde{V}^2}{(A^2+4\tilde{V}^2)^{3/2}}g_{\textbf{k}}^2.\nonumber
\end{eqnarray}

Then we consider $\tilde V$ near the boundary from the side of
hybridization phase where $|\tilde V/A|$ is very small. Here two
situations must be distinguished. When $A>0$, we have
\begin{eqnarray*}
E^{(1)}_{\textbf{k}}&=&A+\frac{{\tilde V}^2}{A}-\mu+\frac{{\tilde
V}^2}{A^2}g_{\textbf{k}}+\frac{{\tilde V}^2}{A^3}g_{\textbf{k}}^2+\cdots\\
E^{(2)}_{\textbf{k}}&=&A+\frac{{\tilde V}^2}{A}-\mu-\frac{{\tilde
V}^2}{A^2}g_{\textbf{k}}+\frac{{\tilde
V}^2}{A^3}g_{\textbf{k}}^2+\cdots\\
E^{(3)}_{\textbf{k}}&=&-\frac{{\tilde V}^2}{A}-\mu+(1-\frac{{\tilde
V}^2}{A^2})g_{\textbf{k}}-\frac{{\tilde V}^2}{A^3}g_{\textbf{k}}^2+\cdots\\
E^{(4)}_{\textbf{k}}&=&-\frac{{\tilde V}^2}{A}-\mu-(1-\frac{{\tilde
V}^2}{A^2})g_{\textbf{k}}-\frac{{\tilde
V}^2}{A^3}g_{\textbf{k}}^2+\cdots.
\end{eqnarray*}

While when $A<0$, we have
\begin{eqnarray*}
E^{(1)}_{\textbf{k}}&=&\frac{{\tilde V}^2}{|A|}-\mu+(1-\frac{{\tilde
V}^2}{A^2})g_{\textbf{k}}+\frac{{\tilde V}^2}{|A|^3}g_{\textbf{k}}^2+\cdots\\
E^{(2)}_{\textbf{k}}&=&\frac{{\tilde V}^2}{|A|}-\mu-(1-\frac{{\tilde
V}^2}{A^2})g_{\textbf{k}}+\frac{{\tilde
V}^2}{|A|^3}g_{\textbf{k}}^2+\cdots\\
E^{(3)}_{\textbf{k}}&=&-|A|-\frac{{\tilde
V}^2}{|A|}-\mu+\frac{{\tilde
V}^2}{A^2}g_{\textbf{k}}-\frac{{\tilde V}^2}{|A|^3}g_{\textbf{k}}^2+\cdots\\
E^{(4)}_{\textbf{k}}&=&-|A|-\frac{{\tilde
V}^2}{|A|}-\mu-\frac{{\tilde V}^2}{A^2}g_{\textbf{k}}-\frac{{\tilde
V}^2}{|A|^3}g_{\textbf{k}}^2+\cdots.
\end{eqnarray*}

Notice that this situation applies to both weak-coupling regime at
the $3/4$-filling and strong-coupling regime at the $1/4$-filling if
we only focus on the Fermi energy close to the band touching points.

By taking into account the trigonal warping effect of the spectrum
near $\textbf{k}=\textbf{k}_D+\textbf{q}$, with
$|\textbf{q}|/|\textbf{k}_D|\ll 1$, we have
\begin{eqnarray}
\frac{|g_{\textbf{q}}|}{t}=\frac{3a_0}{2}|\textbf
q|-\frac{3a_0^2}{8}\sin(3\theta_{\textbf {q}})|\textbf {q}|^2+{\cal
O} (|\textbf{q}|/|\textbf{k}_D|)^3)
\end{eqnarray}
with $\theta_{\textbf{q}}=\arctan(\frac{q_x}{q_y})$. Then Eq.(S4)
and Eq.(S5) are obtained by substituting above expression into
Eq.(S6) for the weak- and strong-coupling regimes respectively.

\section*{In the presence of spin-orbital coupling}

In the presence of the intrinsic SOC, the M-matrix appears in Eq.(2)
becomes
\begin{eqnarray}
M_{\textbf{k}\sigma}=\left(\begin{array}{cc}
                                             \sigma \Lambda _{\textbf{k}}-\mu & \epsilon_{\textbf{k}} \\
                                           \epsilon_{\textbf{k}}^* &  -\sigma \Lambda_{\textbf{k}}-\mu \\
                                         \end{array}
                      \right),
\end{eqnarray}
where,  $\sigma=+1$ and $-1$ refers to spin up and spin down,
$\Lambda_{\textbf{k}}=2\lambda_{so}[\sin{k_1}-\sin{k_2}-\sin{(k_1-k_2)}]=2\lambda_{so}f_{\textbf{k}}$,
$k_1=\textbf{k}\cdot\textbf{a}_1$,
$k_2=\textbf{k}\cdot\textbf{a}_2$. Because the spin-up and spin-down
components are degenerate and the time reversal symmetry is
preserved under $\lambda_{so}$, we only need to consider one of the
component, the spin-up component. So in the following we focus on
the semimetal phase in weak-coupling regime where the two relevant
bands for the spin-up component are

\begin{eqnarray*}
E^{(1)}_{\textbf{k}\uparrow}&=& \frac{1}{2}\{\epsilon_d-\mu+\sqrt{g
_{\textbf{k}}^2+4\lambda^2_{so}f^2_{\textbf{k}}}\\
&&+\sqrt{\left(\epsilon_d+\mu-\sqrt{g_{\textbf{k}}^2+4\lambda^2_{so}f^2_{\textbf{k}}}\right)^2+4{\tilde
V}^2}\},\nonumber\\
E^{(2)}_{\textbf{k}\uparrow}&=& \frac{1}{2}\{\epsilon_d-\mu-\sqrt{g
_{\textbf{k}}^2+4\lambda^2_{so}f^2_{\textbf{k}}}\\
&&+\sqrt{\left(\epsilon_d+\mu+\sqrt{g_{\textbf{k}}^2+4\lambda^2_{so}f^2_{\textbf{k}}}\right)^2+4{\tilde
V}^2}\}.\nonumber
\end{eqnarray*}
At the Dirac points, we have $ \Delta_{12}\equiv
E^{(1)}_{\textbf{k}}-E^{(2)}_{\textbf{k}} =3\sqrt{3}\lambda_{so}
+\frac{1}{2}\{
\sqrt{[\epsilon_d+\mu-3\sqrt{3}\lambda_{so}]^2+4{\tilde V}^2}
-\sqrt{[\epsilon_d+\mu-3\sqrt{3}\lambda_{so}]^2+4{\tilde V}^2}\}$.
When $\lambda_{so}$ is very small,
$\Delta_{12}=3\sqrt{3}\lambda_{so}(1-\frac{|\epsilon_d+\mu|}{\Delta})+{\cal
O}(\lambda^2_{so})$.

Therefore, a bulk band gap opens immediately by turning on the SOC.
Obviously, the similar conclusion applies to the strong-coupling
regime at the $1/4$-filling. Because in the absence of the SOC the
semimetal phase is topological non-trivial owing the $\pi$-Berry
phase around the Dirac points, gap opening which does not destroy
the time reversal symmetry will lead to non-vanishing flow of the
Berry curvature from the band $E^{(1)}_{\textbf{k}\uparrow}$ to the
band $E^{(2)}_{\textbf{k}\uparrow}$ or vice verse, from the band
$E^{(2)}_{\textbf{k}\downarrow}$ to the band
$E^{(1)}_{\textbf{k}\downarrow}$.  This will result in the edge
states protected by the time reversal symmetry. The spectra obtained
for a finite strip system with the zig-zag geometry in the
strong-coupling regime at the $1/4$-filling do exhibit the surface
states as shown in Fig.(2).

\end{document}